\documentclass[Journal,twocolumn]{IEEEtran}

\IEEEoverridecommandlockouts

\usepackage{color}
\usepackage{graphicx}
\usepackage{amsmath}
\usepackage{amssymb}
\usepackage{algorithm}
\usepackage{algorithmic}
\usepackage{amsmath}
\usepackage{multirow}
\usepackage{booktabs}
\usepackage{array}
\usepackage{amsthm}
\usepackage{stfloats}
\usepackage{caption}
\usepackage{bm}
\usepackage{epstopdf}
\usepackage{gensymb} 
\usepackage{subfigure} 
\usepackage{url} 
\usepackage{flushend}

\newcommand{\be}{\begin{equation}}
	\newcommand{\ee}{\end{equation}}
\newcommand{\bea}{\begin{eqnarray}}
	\newcommand{\eea}{\end{eqnarray}}
\newcommand{\ba}{\begin{array}}
	\newcommand{\ea}{\end{array}}

\makeatletter

\newcommand{\Rmnum}[1]{\expandafter\@slowromancap\romannumeral #1@}
\makeatother

\newcommand{\RNum}[1]{\uppercase\expandafter{\romannumeral #1\relax}}

\title{Target Detection in OFDM-ISAC Systems:\\ A Multipath Exploitation Approach 
\thanks{X. Lv, and M. Li are with the School of Information and Communication Engineering, Dalian University of Technology, Dalian 116024, China (e-mail: lvxiaohan@mail.dlut.edu.cn; mli@dlut.edu.cn).}
\thanks{R. Liu is also with the Center for Pervasive Communications and Computing, University of California, Irvine, CA 92697, USA (e-mail: rangl2@uci.edu).}
\thanks{Q. Liu is with the School of Computer Science and Technology, Dalian University of Technology, Dalian 116024, China (e-mail: qianliu@dlut.edu.cn).}
}

\author{Xiaohan~Lv,
        Rang~Liu,~\IEEEmembership{Member,~IEEE,}
        Ming~Li,~\IEEEmembership{Senior~Member,~IEEE,}
        and~Qian~Liu,~\IEEEmembership{Member,~IEEE}
}

\begin{document}
\maketitle
\pagestyle{empty}
\thispagestyle{empty}
	
\vspace{-0 cm}
\begin{abstract}
This paper investigates the potential of multipath exploitation for enhancing target detection in orthogonal frequency division multiplexing (OFDM)-based integrated sensing and communication (ISAC) systems. The study aims to improve target detection performance by harnessing the diversity gain in the delay-Doppler domain. We propose a weighted generalized likelihood ratio test (GLRT) detector that effectively leverages the multipath propagation between the base station (BS) and the target. To further enhance detection accuracy, a joint optimization framework is developed for subcarrier power allocation at the transmitter and weight coefficients of the GLRT detector. The objective is to maximize the probability of target detection while satisfying constraints on total transmit power and the communication receiver’s signal-to-noise ratio (SNR). An iterative algorithm based on the majorization-minimization (MM) method is employed to address the resulting non-convex optimization problem. Simulation results demonstrate the efficacy of the proposed algorithm and confirm the benefits of multipath exploitation for target detection in OFDM-ISAC systems under multipath-rich environments.
\end{abstract}
	
	
\section{Introduction}
Integrated sensing and communication (ISAC) has emerged as a key enabling technology for sixth-generation (6G) networks \cite{ISAC summary}-\cite{sensing assisted comm}. As ISAC systems are typically deployed in complex urban environments, they must contend with the challenges posed by multipath wireless propagation. While traditional communication systems mitigate multipath effects using techniques such as equalization and orthogonal frequency division multiplexing (OFDM), radar systems are often designed for line-of-sight (LoS) conditions, as multipath echoes can introduce ambiguities and degrade sensing performance. Consequently, handling multipath propagation remains a critical challenge for radar applications. 

In recent years, there has been growing interest in exploiting multipath propagation to enhance radar sensing capabilities. For instance, the work in \cite{diss1} investigated the use of multipath and sparse reconstruction techniques in a distributed multi-static radar network. Similarly, \cite{diss2} explored the identification of moving objects in OFDM radar systems aided by multipath signals. The potential of specular reflection multipath under incomplete information was studied in \cite{diss3}. Furthermore, \cite{diss4} proposed the use of an active reconfigurable intelligent surface (RIS) to create additional paths, enabling improved observation of targets illuminated by a radar transmitter.

This paper aims to explore multipath exploitation for target detection in OFDM-ISAC systems by leveraging the inherent diversity of multipath propagation. Specifically, we develop a comprehensive model of multipath channels in OFDM-ISAC systems, capturing essential characteristics such as distinct time delays and Doppler shifts. To enhance detection performance, we propose a weighted generalized likelihood ratio test (GLRT) detector that effectively utilizes the delay-Doppler properties of the different propagation paths.
To further improve detection accuracy, we formulate a joint optimization problem involving the transmit power allocation across subcarriers and the weight coefficients of the GLRT detector. The objective is to maximize the target detection probability while satisfying constraints on the total transmit power and communication signal-to-noise ratio (SNR). The resulting non-convex problem is addressed using an alternating optimization approach based on the majorization-minimization (MM) framework. Simulation results validate the proposed algorithm, demonstrating the benefits of multipath exploitation for target detection in OFDM-ISAC systems. The contributions of this work go beyond algorithm development, encompassing a rigorous system model and an effective joint design that fully leverages multipath diversity to enhance performance.
 
\begin{figure}[t]
\centering
\includegraphics[width= 2.8in]{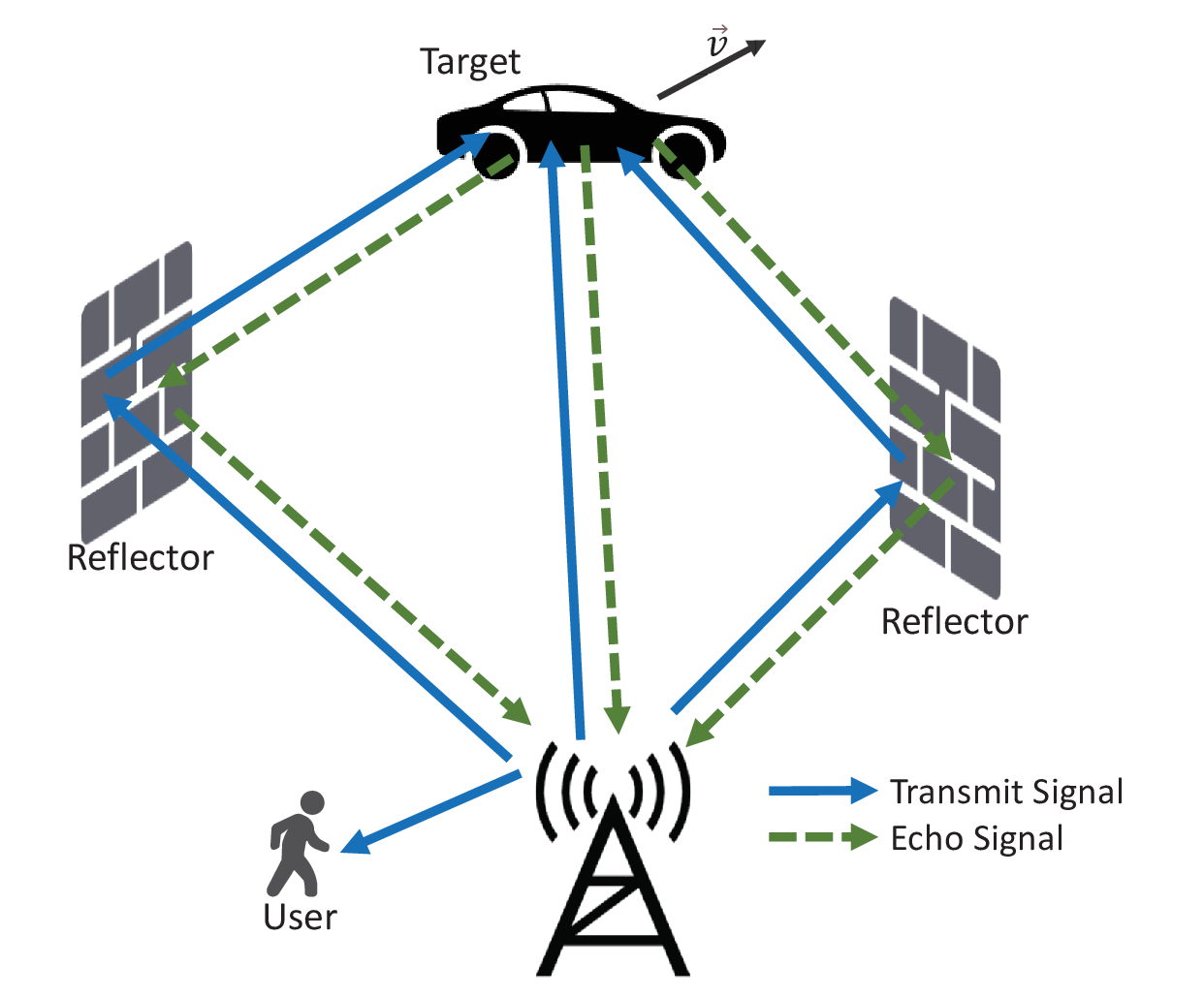}\\
\caption{An ISAC system under multipath propagation scenario.}\label{fig:system model}\vspace{0 cm}
\end{figure}
 
\section{System Model}
We consider an ISAC system deployed in a multipath environment as shown in Fig. \ref{fig:system model}. Specifically, the BS utilizes the dual-functional OFDM waveform to simultaneously communicate with a single-antenna communication user and detect a potential target. 
We assume the total bandwidth $B$ is divided into $N$ subcarriers, and the corresponding bandwidth spacing and symbol duration are denoted as  $\Delta f = B/N$ and $T = 1/\Delta f$, respectively. Moreover, a frame of $M$ OFDM symbols is utilized for radar sensing, encompassing precise estimations of both range and velocity. By applying an $N$-point inverse fast Fourier transform (IFFT) at the transmitter, we can express the time domain OFDM-ISAC signal within a single frame as follows
\begin{equation} \label{eq:transmit signal} 
s(t)=\sum_{m=1}^{M}\sum_{n=1}^{N}{a}_{n,m}{x}_{n,m} g_{\mathrm{ tx}}(t-mT) e^{\jmath2\pi (f_0 + n \Delta f(t-mT))}, 
\end{equation}
in which ${x}_{n,m}$ represents the communication symbol transmitted on the $n$-th subcarrier within the $m$-th symbol-slot, ${a}_{n,m}$ refers to the amplification factor to ${x}_{n,m}$, $g_{\mathrm{tx}}(t-mT)$ denotes the pulse shaping filter, and $f_0$ is the central carrier frequency.

At the communication receiver side, after down-conversion, cyclic prefix (CP) removal and $N$-point fast Fourier transform (FFT), the received baseband signal on the $n$-th subcarrier during the $m$-th symbol-slot is obtained as 
\begin{equation}\label{eq:user receive signal}{y}_{\mathrm{c},n,m}=h_{{\mathrm{c}},{{n}}}{a}_{n,m}{x}_{n,m}+{z}_{\mathrm{c,}n,m},
\end{equation}
where $h_{{\mathrm{c}},{{n}}}$ denotes the frequency domain communication channel between the BS and the user, and ${z}_{\mathrm{c,}n,m}\sim \mathcal{C}\mathcal{N}(0,\sigma_\mathrm{\mathrm{c}}^{2})$ represents the additive white Gaussian noise (AWGN). Based on this model, the communication SNR on the $n$-th subcarrier during the $m$-th symbol-slot can be written as
\begin{equation}\label{SNR of comm}
\mathrm{SNR}_{\mathrm{c,} n,m}=\frac{\vert h_{\mathrm{c},{n}}{a}_{n,m}{x}_{n,m}\vert^2}{\sigma_\mathrm{c}^2},~\forall n,m.
\end{equation}

Although OFDM has been widely adopted in wireless communications, its application in ISAC systems—particularly in multipath propagation environments—remains underexplored. This study seeks to harness the diversity gain provided by multipath propagation to enhance target detection performance in OFDM-ISAC systems. To this end, we first establish a comprehensive multipath channel model for radar sensing, capturing the diversity inherent in the delay-Doppler domain.

Without loss of generality, we assume that there are $L$ distinct propagation paths within the BS-target-BS round-trip channel, which exhibit distinguishable time delays and Doppler shifts. 
For the $l$-th path, $l=1, 2, \ldots, L$, the channel impulse response, characterized by a specific delay $\tau_l$ and a Doppler shift $\nu_l$, can be concisely represented as \cite{otfs}
\begin{equation}\label{eq:hdd_l}
h_l(\displaystyle \tau,\nu)=\beta_{l}\delta(\tau-\tau_{l})\delta(\nu-\nu_{l}),
\end{equation}
where $\beta_{l}$ denotes the path loss, and $\bm \beta\triangleq[\beta_1,\cdots,\beta_l]$. It is obvious that the channels $h_l(\displaystyle \tau,\nu)$ exhibit favorable orthogonality for different parameters $\tau_l$ and $\nu_l$. 
In discrete-time models, delay and Doppler shifts are typically represented as integer multiples of their respective resolutions. Specifically, the discretized delay shift is expressed as \( \hat{\tau}_l = k_l / (N \Delta f) \), where \( k_l = \mathrm{round}[\nu_l MT] \) denotes the delay tap index. Similarly, the discretized Doppler shift is given by \( \hat{\nu}_l = r_l / (MT) \), with \( r_l = \mathrm{round}[\tau_l N \Delta f] \) representing the Doppler tap index.
Accordingly, the channel for the \(l\)-th propagation path in the delay-Doppler domain can be re-formulated as   
\begin{equation}\label{eq:hdd_l}
h_l(\displaystyle \tau,\nu)=\beta_{l}\delta(\tau-\hat{\tau}_l)\delta(\nu-\hat{\nu}_l).
\end{equation}

Given the complex baseband channel impulse response in the delay-Doppler domain, we transform the channel model into the time-frequency domain. To achieve this, the symplectic finite Fourier transform (SFFT) is employed to map the delay-Doppler channel into its time-frequency representation. 
In general, time-frequency domain samples are positioned at integer multiples of the symbol duration $T$ and the subcarrier spacing $\Delta f$. Accordingly, the channel coefficient for the $l$-th path on the $n$-th subcarrier within the $m$-th symbol-slot is expressed as
\begin{equation}\label{eq:h_nml of sensing}\begin{aligned}  
h_{n,m,l} &=\int_\nu\int_\tau h_l(\tau,\nu) {e^{\jmath 2\pi \nu nT}} e^{-\jmath 2\pi (\nu + m\Delta f)\tau } d\tau d\nu\\
&= \beta_l e^{\jmath 2\pi \hat{\nu}_lnT}e^{-\jmath2\pi(\hat{\nu}_l+m\Delta f)\hat{\tau}_l}. 
\end{aligned}\end{equation}

The transmitted dual-functional signal propagates to the target and reflects back to the BS through 
$L$ distinct channels. At the BS receiver, following down-conversion, CP removal, and 
$N$-point fast Fourier transform (FFT), the baseband echo signal on the $n$-th subcarrier during the $m$-th symbol-slot is expressed as \begin{equation}\label{eq:receive ecch ynm}{y}_{n,m}=\sum_{l=1}^{L}\lambda _{n,l} {h}_{n,m,l}{a}_{n,m}{x}_{n,m}+{z}_{n,m}, \end{equation}
where $\lambda _{n,l}\sim\mathcal{CN}(0,\sigma^2_{n,l})$ denotes  the frequency-dependent radar cross section (RCS) for the $l$-th path, and $z_{n,m}\sim\mathcal{CN}(0,\sigma_\text{r}^2)$ represents the AWGN.

For algorithm development, we stack the received signals across all subcarriers for the 
$m$-th symbol-slot and represent the received signal in vector form as 
\begin{equation}\label{eq:receive echo ym}\mathbf{y}_{m}=\sum_{l=1}^{L}\boldsymbol\Lambda _{l}{\mathbf{H}}_{m,l}\mathbf{A}_{m}\mathbf{x}_{m}+\mathbf{z}_m,\end{equation}
where we define the received signals, the communication symbols, the amplifications, and the noise during the $m$-th symbol-slot as $\mathbf{y}_{m}\triangleq[y_{1,m}, y_{2,m}, \ldots, y_{N,m}]^T$, $\mathbf{x}_{m} \triangleq [x_{1,m}, x_{2,m},\ldots, x_{N,m} ]^T$, $\mathbf{a}_m\triangleq [a_{1,m}, a_{2,m}, \ldots, a_{N,m}]^T$, and $\mathbf{z}_{m} \triangleq [z_{1,m}, z_{2,m}, \ldots, z_{N,m}]^T \sim\mathcal{CN}(\mathbf{0},{\bm\Sigma})$, respectively. The $l$-th channel impulse response during the $m$-th symbol-slot is defined as $\mathbf{h}_{m,l} \triangleq [h_{1,m,l}, h_{2,m,l}, \ldots, h_{N,m,l}]^T$,  with its corresponding channel matrix given by $\mathbf{H}_{m,l} \triangleq \mathrm{diag}\{\mathbf{h}_{m,l}\}$. Additionally,  $\boldsymbol{\Lambda}_l \triangleq \mathrm{diag}\{\lambda_{1,l}, \lambda_{2,l}, \cdots, \lambda_{N,l}\}$  is the RCS matrix, and $\mathbf{A}_m \triangleq \mathrm{diag}\{\mathbf{a}_m\}$ is the amplification matrix.

Next, we collect the received signals across all 
$M$ symbol-slots and express them as
\be \label{eq:Ytf}
\mathbf{Y} \triangleq[\mathbf{y}_{1},\mathbf{y}_{2},\cdots,\mathbf{y}_{M} ]=\sum\limits_{l = 1}^{L} \boldsymbol\Lambda _l\mathbf{H}_{l}\mathbf{AX}+\mathbf{Z},
\ee
where we define the $l$-th channel matrix ${\mathbf{H}}_l \triangleq [ {\mathbf{H}}_{1,l}, \mathbf{H}_{2,l},\ldots, {\mathbf{H}}_{M,l} ]$, the power allocation matrix $\mathbf{A}\triangleq\mathrm{diag} \{ {\mathbf{a}}\}$ with $\mathbf{a}\triangleq [\mathbf{a}_1^T,\mathbf{a}_2^T,\ldots,\mathbf{a}_M^T]^T$, the communication symbol matrix $\mathbf{X}\triangleq\mathrm{blkdiag}\big \{ {\mathbf{x}}_{1}, \mathbf{x}_2,\ldots, {\mathbf{x}}_{M}\big \}$, and the noise matrix $\mathbf{Z}\triangleq[\mathbf{z}_{1},\mathbf{z}_{2},\cdots,\mathbf{z}_{M}]$, respectively.

\section{GLRT-based Target Detector}
For target detection in the considered OFDM-ISAC system, our objective is to utilize the echo signal in \eqref{eq:Ytf} to determine the presence of a target within the specified range cell and  delay-Doppler bin. This detection task can be formulated as the following binary hypothesis test:
\begin{eqnarray} \label{hypotheses}
	\left\{\begin{array}{l}
		{\cal H}_0: {\mathbf{Y}}=\mathbf{Z}, \\
		{\cal H}_1: {\mathbf{Y}}= \sum\limits_{l = 1}^{L}\boldsymbol\Lambda _l\mathbf{H}_{l}\mathbf{AX}+\mathbf{Z},
	\end{array}\right.
\end{eqnarray}
where ${\mathcal H_0}$ and ${\mathcal H_1}$ represent hypotheses for the absence and presence of the target, respectively. 
Without prior knowledge of the RCS reflection coefficient $\boldsymbol\Lambda_l$ and noise covariance matrix ${\bm\Sigma}$, we adopt the widely used  GLRT detector for target detection. Following the Neyman-Pearson detector \cite{np detector}, we can formulate the probability of detection for a given probability of false alarm. The GLRT detector is expressed as 
\begin{align}\label{eq:GLRT by lf}
\eta  = \frac{{\mathop {\max }\limits_{{\hat{\boldsymbol\Lambda}_l},{ \hat{\bm\Sigma}}}~~f_1({\mathbf Y} \mid_{\mathbf{H}_{l},\mathbf{A,X}, {\hat{\boldsymbol\Lambda}_l} ,\hat{\bm\Sigma}} ;\mathcal{H}_1)}}{{\mathop {\max }\limits_{{ \hat{\bm\Sigma}}}~~f_0({\mathbf Y} \mid_{{ \hat{\bm\Sigma}}};{\mathcal{H}}_0)}} \underset{{\cal H}_0}{\overset{{\cal H}_1}{\gtrless} \xi},
\end{align}
where $f_0$ and $f_1$ are the likelihood functions under ${\mathcal H_0}$ and ${\mathcal H_1}$, respectively, and $\xi$ is the detection threshold. The matrix  ${\hat{\bm\Sigma}}$ is the maximum likelihood estimation (MLE) of $\bm\Sigma$, and the MLE of $\boldsymbol\Lambda_l$ can be easily calculated by ${\hat{\boldsymbol\Lambda}_l} \triangleq\mathbf{P}_{\mathrm{s},l}\mathbf{Y}\mathbf{X}^{\dagger}\mathbf{A}^{\dagger}\mathbf{H}_{l}^{\dagger}$, where the projection matrix for the channel ${\widetilde{\mathbf{H}}_{l}}\triangleq[\mathbf{h}_{1,l},\mathbf{h}_{2,l},\cdots,\mathbf{h}_{M,l}]$
is defined as 
$\mathbf{P}_{\mathrm{s}, l}\triangleq{{\widetilde{\mathbf{H}}_{l}}({{\widetilde{\mathbf{H}}_{l}}^H}{{\widetilde{\mathbf{H}}_{l}}})^{\dagger}{{\widetilde{\mathbf{H}}_{l}}}^H}$. 
Due to the orthogonality between the channels of different paths, i.e., $\widetilde{\mathbf H}_{l_1} \widetilde{\mathbf H}_{l_2}^H = \mathbf 0, \forall l_1 \neq l_2 $, the channel projection matrices are also orthogonal to each other, i.e.,  $\mathbf P_{\text{s},l_1}\mathbf P_{\text{s},l_2}^H =\bm 0, \forall l_1 \neq l_2$. In addition, it is also noted that $\widetilde{\mathbf H}_{l}$ and $\mathbf P_{\text{s},l}$ are rank-one matrices.

The matrix 
${\mathbf Y}$ follows a zero-mean Gaussian distribution Under ${\mathcal{H}}_0$, while under ${\mathcal{H}}_1$ it follows a Gaussian distribution with mean 
$\sum_{l = 1}^{L}{{\boldsymbol\Lambda}_l}\mathbf{H}_{l}\mathbf{AX}$. The GLRT detector can therefore be reformulated as \cite{glrt rate}: 
\begin{align}\label{eq:GLRT by Y}
\eta  &= \frac{\lVert{\mathbf Y}\rVert_F^2 }{\lVert{\mathbf Y} - \sum_{l = 1}^{L}{\hat{\boldsymbol\Lambda}_l}\mathbf{H}_{l}\mathbf{AX}\rVert_F^2} \notag\\&= \frac{\lVert{\mathbf Y}\rVert_F^2 }{\lVert{\mathbf Y} - \sum_{l = 1}^{L}\mathbf{P}_{\mathrm{s}{,l}}\mathbf Y \rVert_F^2} \underset{{\cal H}_0}{\overset{{\cal H}_1}{\gtrless} \xi  }.
\end{align}
The orthogonality of the projection matrices $\mathbf{P}_{\mathrm{s}{,l}},~\forall l$  allows us to apply the Craig-Sakamoto theorem \cite{cs theorem}, leading to the following compact form of the GLRT detector:
\begin{align}\label{eq:GLRT with multipath}
\eta & = 1+ \frac{\Vert\mathbf P_{\mathrm{s},1} \mathbf Y \Vert^2_F+\cdots+  \Vert \mathbf P_{\mathrm{s},L} \mathbf Y \Vert^2_F}{\Vert \mathbf P_\mathrm{n}\mathbf Y \Vert^2_F} \underset{{\cal H}_0}{\overset{{\cal H}_1}{\gtrless} \xi  } ,
\end{align}
where the null space of the overall projection matrix $\mathbf{P}_\mathrm{s} \triangleq \sum_{l=1}^{L}\mathbf{P}_{\mathrm{s},l}$ is defined as 
$\mathbf{P}_\mathrm{n} \triangleq \mathbf{I}_{MN}-\mathbf{P}_\mathrm{s}$.

In multipath scenarios, we aim to enhance detection performance by leveraging the diverse reception of echoes from multiple propagation paths. To this end, we propose a weighted GLRT detector, modifying (\ref{eq:GLRT with multipath}) as follows:  
\begin{equation}
\label{eq:GLRT weight}
\widetilde{\eta} = 1+ w_1^2\frac{\Vert\mathbf P_{\mathrm{s},1} \mathbf Y \Vert^2_F}{\Vert \mathbf P_\mathrm{n} \mathbf Y\Vert^2_F}+\cdots+w_L^2\frac{\Vert\mathbf P_{\mathrm{s},L} \mathbf Y \Vert^2_F}{\Vert \mathbf P_\mathrm{n} \mathbf Y\Vert^2_F} \underset{{\cal H}_0}{\overset{{\cal H}_1}{\gtrless} \xi} ,
\end{equation}
where $w_l$, $l=1, 2, \ldots, L$ are the weight coefficients for each propagation path, $\sum_{l=1}^L w_l^2 = 1$. 
Using the proposed weighted GLRT detector, we analyze the probability of false alarm under ${\mathcal H_0}$ and the probability of detection under ${\mathcal H_1}$ as follows.

Under \(\mathcal{H}_0\), the matrix \(\mathbf{Y}\) follows a complex normal distribution, expressed as \(\mathbf{Y} \sim \mathcal{CN}(\mathbf{0}, \mathbf{I}_M \otimes \boldsymbol{\Sigma})\). Owing to the orthogonality of the projection matrices \(\mathbf{P}_{\mathrm{s},l}\), the proposed detector in (\ref{eq:GLRT weight}) can be reformulated under \(\mathcal{H}_0\) as  
\begin{equation} \label{eq:H10 GLRT result}  
\widetilde{\eta}_0 = 1 + \frac{\left\| w_1 \mathbf{P}_{\text{s},1} \mathbf{Z} + \cdots + w_L \mathbf{P}_{\text{s},L} \mathbf{Z} \right\|_F^2}{\left\| \mathbf{P}_{\text{n}} \mathbf{Z} \right\|_F^2} \underset{\mathcal{H}_0}{\overset{\mathcal{H}_1}{\gtrless}} \xi.  
\end{equation}   
By leveraging the independence of orthogonal projection matrices and the linear transformation properties of Gaussian distributions, we obtain  
$\sum_{l=1}^{L} w_l \mathbf{P}_{\text{s},l} \mathbf{Z} \sim \mathcal{CN} ( \mathbf{0}, \sum_{l=1}^{L} w_l^2 (\mathbf{I}_M \otimes\mathbf{P}_{\text{s},l}  \bm\Sigma \mathbf{P}_{\text{s},l}^H))$.  
Based on sampling theory,, the detection statistic \(\widetilde{\eta}_0\) under \(\mathcal{H}_0\) follows an F-distribution.
When the weight coefficient vector \(\mathbf{w} \triangleq [w_1, w_2, \ldots, w_L]^T\) is normalized such that \(\|\mathbf{w}\|^2 = 1\), the resulting F-distribution depends only on \(N\), \(M\), and \(L\). In other words, the design of the power allocation matrix \(\mathbf{A}\) and the weight coefficient vector \(\mathbf{w}\) does not influence the distribution of \(\widetilde{\eta}_0\), ensuring that the probability of false alarm remains unchanged.

Under ${\mathcal H_1}$, the observation $\mathbf{Y}$ follows a complex normal distribution ${\mathbf Y}\sim{\cal CN}({\sum_{l = 1}^{L}\boldsymbol\Lambda_l\mathbf{H}_{l}\mathbf{AX}},~{\mathbf I}_{M}\otimes{\bm\Sigma})$. Thus, the proposed weighted GLRT detector is expressed as
\be \hspace{-0.2 cm} \label{eq:H1 GLRT result}\widetilde{\eta}_1 =1+ \frac{\Vert \sum_{l=1}^{L}w_l\mathbf P_{\text{s},l}(\sum_{l = 1}^{L}\boldsymbol\Lambda _l\mathbf{H}_{l}\mathbf{AX}+\mathbf{Z}) \Vert^2_F}{\Vert \mathbf P_\text{n}(\sum_{l = 1}^{L}\boldsymbol\Lambda _l\mathbf{H}_{l}\mathbf{AX}+\mathbf{Z}) \Vert^2_F}\underset{{\cal H}_0}{\overset{{\cal H}_1}{\gtrless} \xi  }.\ee
The output $\widetilde{\eta}_1$ in (\ref{eq:H1 GLRT result}) follows a non-central F-distribution, with the non-centrality parameter given by  $\Vert \sum_{l = 1}^{L} w_l{\hat{\boldsymbol\Lambda} _l\mathbf{H}_{l}\mathbf{AX}} \Vert^2_F= {\sum_{l = 1}^{L} w_l^2\Vert {\hat{\boldsymbol\Lambda} _l\mathbf{H}_{l}\mathbf{AX}} \Vert^2_F }$. This non-centrality parameter is proportional to the signal strength and, consequently, to the detection probability \cite{diss2}.
To improve detection performance, we aim to maximize the non-centrality parameter $ {\sum_{l = 1}^{L} w_l^2\Vert {\hat{\boldsymbol\Lambda} _l\mathbf{H}_{l}\mathbf{AX}} \Vert^2_F }$. However, directly estimating the RCS matrix ${\boldsymbol\Lambda} _l$ can be computationally demanding or infeasible.  To address this challenge, we propose maximizing the non-central parameter without explicitly estimating ${\boldsymbol\Lambda} _l$. Specifically, the objective function is formulated as 
\be \label{eq:objective function0}
		f(\mathbf {A},\mathbf {w})\triangleq  {\sum\limits_{l = 1}^{L} w_l^2\Vert {\mathbf{H}_{l}\mathbf{AX}} \Vert^2_F } .
\ee
The value of $f(\mathbf {A},\mathbf {w})$ is positively proportional to the detection probability.Therefore, to improve target detection performance, we investigate the joint design of power allocation matrix  $\mathbf{A}$ and weight coefficient  vector $\mathbf{w}$ by maximizing $f(\mathbf {A},\mathbf {w})$ under both the communication SNR requirement and the power budget constraints.

\section{Joint Transmit Power Allocation and Detector Weight Coefficient Design}
Based on the proposed weighted GLRT detector \eqref{eq:GLRT weight}, we aim to maximize the detection probability by jointly designing the power allocation matrix $\mathbf{A}$ for the transmitter and weight vector $\mathbf{w}$ for the detector. 
This optimization problem can be formulated as
\begin{subequations}\begin{align} \max _{\mathbf{A} ,\mathbf {w}}  \;\;&f(\mathbf {A},\mathbf {w}) \label{objective function a}\\ \mathrm{s.t.} \quad & \Vert \mathbf {A}\Vert ^{2}_{F} \leq P_{\mathrm{BS}}, \label{objective function b}\\ & {\vert h_{\mathrm{c},{n}}{a}_{n,m}{x}_{n,m}\vert^2}/{\sigma_\mathrm{c}^2} \geq \gamma, ~\forall n,m, \label{objective function c} \\
& \| \mathbf{w}\|^{2} = 1,  \label{objective function d} 
\end{align} \end{subequations}
where (\ref{objective function b}) imposes transmit power constraint and $P_{\mathrm{BS}}$ is power budget, and (\ref{objective function c}) is the SNR requirement of the communication user. This optimization problem is highly non-convex due to the complex multivariate objective function in \eqref{objective function a} and the coupling between variables. To overcome these challenges, we decompose the original problem into two tractable sub-problems, which are solved iteratively in an alternating manner.

\subsection{Power Allocation Design for Transmitter} 
In this subsection, we present the BS power allocation design. Given the fixed weight coefficient $\mathbf{w}$, the optimization problem can be simplified as
\begin{align}\label{objective function to A}
\max _{\mathbf {a}} \;\;& \sum\limits_{l = 1}^{L} w_l^2\Vert {\mathbf{H}_{l}\mathbf{AX}} \Vert^2_F  \\ \mathrm{s.t.} \;\;\;& \mathrm{(\ref{objective function b}),~(\ref{objective function c})}. \notag
\end{align}
To facilitate algorithm development, we leverage the fact that $\mathbf{A}$ is diagonal, while 
$\mathbf{X}$ and $\mathbf{H}_l$ are block-diagonal. This allows us to reformulate    (\ref{objective function to A}) to an equivalent form by
\begin{align}\label{eq:derive aRa}  
\sum\limits_{l = 1}^{L} w_l^2\Vert {\mathbf{H}_{l}\mathbf{AX}} \Vert^2_F &=\sum_{l = 1}^{L}w_l^2 \sum\limits_{m = 1}^{M}\Vert \mathbf{H}_{l}\mathbf{A}\mathbf{s}_m\Vert^2_2 \notag\\	& = \sum_{l = 1}^{L}w_l^2 \sum\limits_{m = 1}^{M}\Vert \mathbf{H}_{l}\mathrm{diag}(\mathbf{s}_m)\mathbf{a}\Vert^2_2 \notag\\ & =  \mathbf{a}^H\mathbf{R}\mathbf{a},	
\end{align}
where for brevity we define $\mathbf{s}_m \triangleq \mathbf{e}_m \otimes \mathbf{x}_m$ and $\mathbf {R} \triangleq \sum_{l = 1}^{L}\sum_{m = 1}^{M}w_l^2 \mathrm{diag}(\mathbf{s}_m^H)\mathbf{H}_l^H\mathbf{H}_l \mathrm{diag}(\mathbf{s}_m)$.
Now the objective function is now expressed as a quadratic form with respect to the power allocation vector $\mathbf{a}$. However, since this quadratic objective is non-concave, we employ the MM method to iteratively approximate it with a sequence of concave surrogate functions. 
Specifically, upon having the solution $\widetilde{\mathbf {a}}_k$ in the $k$-th iteration, we construct a surrogate function that closely approximates the original objective at the current point, serving as a lower bound for maximization in the next step. 
Using the first-order Taylor expansion, the surrogate function for $\mathbf{a}^H\mathbf{R}\mathbf{a}$ can be expressed as
\begin{equation}\label{eq:Taylor expansion} \mathbf {a}^{H}\mathbf {R}\mathbf {a} \geq \widetilde{\mathbf {a}}_{k}^{H}\mathbf {R}\widetilde{\mathbf {a}}_{k} + 2\Re \left\lbrace \widetilde{\mathbf {a}}_{k}^{H}\mathbf {R}\left(\mathbf {a}-\widetilde{\mathbf {a}}_{k}\right) \right\rbrace. \end{equation}
While the power constraint (\ref{objective function b}) is clearly convex, the SNR requirement (\ref{objective function c}) is non-convex in its original form. To address this issue, we use basic algebraic transformations to convert this non-convex constraint into a linear constraint, as demonstrated in (\ref{f:MM result}c). Consequently, the transmit power allocation design problem at point $\widetilde{\mathbf {a}}_{k}$ can be formulated as
\begin{subequations}\label{f:MM result}\begin{align}\max _{\mathbf {a}} \;\;& \Re \left\lbrace \widetilde{\mathbf {a}}_{k}^{H}\mathbf {R}\mathbf {a} \right\rbrace \\ \mathrm{s.t.} \quad & \Vert \mathbf {A}\Vert ^{2}_{F} \leq P_{\mathrm{BS}},  \\ & a_{n,m}|h_{\mathrm{c},{n}}{x}_{n,m}| \geq \sqrt {\gamma/\sigma^2_\text{c}}, ~\forall n,m,  	\end{align}\end{subequations}
which is a convex problem and can be readily solved.

\subsection{Weight Coefficient Design for Detector}
Given the power allocation vector $\mathbf{a}$, the optimization problem of solving for $\mathbf{w}$ can be expressed as 
\begin{subequations}\label{eq:solve for w}\begin{align} \max _{\mathbf{w}}  \;\;&{\sum\limits_{l = 1}^{L} w_l^2\Vert {\mathbf{H}_{l}\mathbf{AX}} \Vert^2_F } = |\mathbf{w}^H\mathbf{d}|^2\\ \mathrm{s.t.} \quad & \| \mathbf{w}\|^{2} = 1, 
\end{align} \end{subequations}
where we exploit the orthogonality between $\mathbf{H}_l,~\forall l$, and define $\mathbf{d} \triangleq [\|\mathbf{H}_1\mathbf{AX}\|_F,\|\mathbf{H}_2\mathbf{AX}\|_F,\ldots,\|\mathbf{H}_L\mathbf{AX}\|_F]^T$. It is clear that problem (\ref{eq:solve for w}) is a typical Rayleigh quotient problem, which can be easily solved by using the Lagrange multiplier method. The optimal solution to $\mathbf{w}$ is given by
 \begin{equation}\label{eq:w_l} \mathbf{w} = {\mathbf{d}}/{\|\mathbf{d}\|_2}.\end{equation}

In summary, the proposed joint power allocation and detector weight coefficient design algorithm is straightforward. We alternatively update $\mathbf{a}$ by solving problem (\ref{f:MM result}) and $\mathbf{w}$ by (\ref{eq:w_l}) until convergence.

\section{Simulation Results}
\begin{table}[!t]
	\centering
	\caption{EXPERIMENTAL CONFIGURATION} \label{table:one}
	\begin{tabular}{|c|c|c|c|}
		\hline
		Parameters  & Values & Parameters  & Value \\
		\hline
		$M$  & 64 & BS location & (0m, 0m)  \\
		\hline
		$N$ & 64 & Reflector location & (-30m, 10m), (20m, 30m)\\
		\hline
		$L$ & 6 & Target location & (0m, 55m) \\
		\hline
		$P_{\mathrm{BS}}$ & 20dBW & Target velocity & (30m/s, 50m/s) \\
		\hline
		$\gamma$ & 8dB & \multirow{2}{*}{$\bm \beta$ } & 9.84 1.02 1.61   \\
		\cline{1-2}
		$\sigma_\text{c}^2$, $\sigma_\text{r}^2$ & -80dBm   &    & 0.10 0.17 0.26\\
		\hline
	\end{tabular}
\end{table}

\begin{figure}[!t]
	\centering
	\includegraphics[width= 2.8 in ]{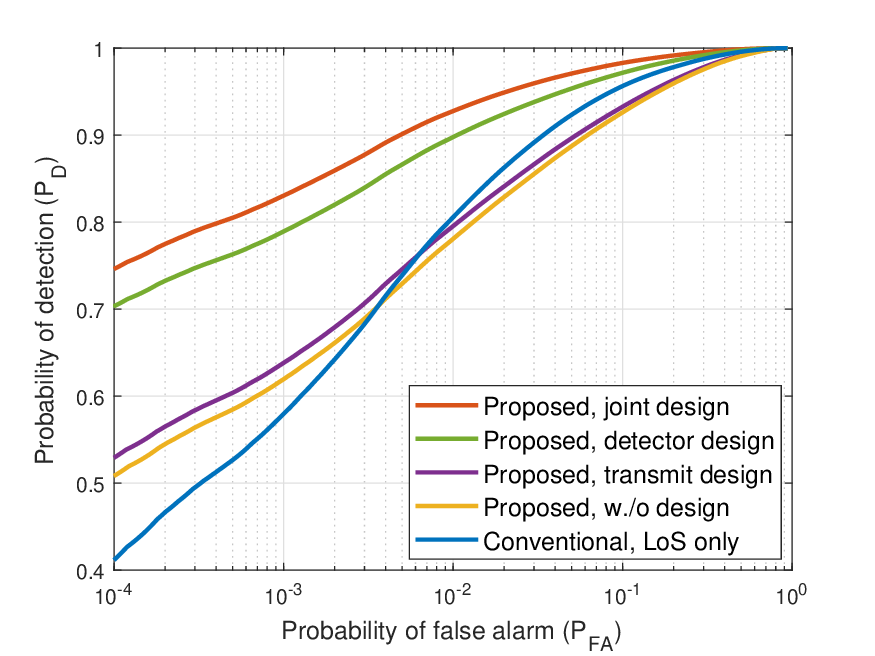}\\
	\caption{Probability of detection versus probability of false alarm.}\label{fig:ROC1} \vspace{-0.2 cm}
 
	\includegraphics[width= 2.8 in]{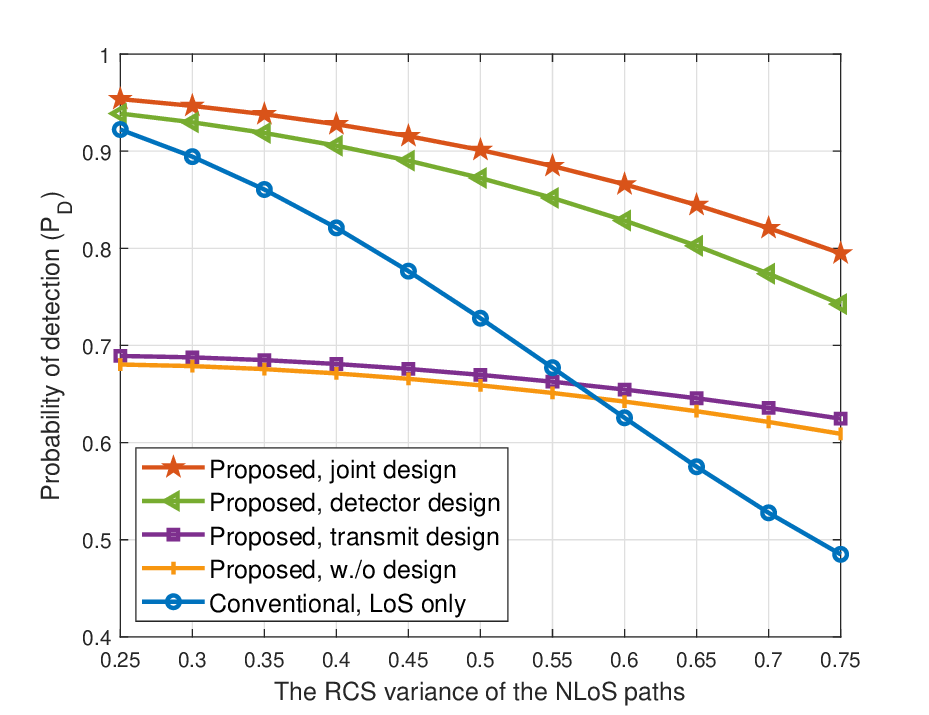}\\
	\caption{Probability of detection versus the RCS variance of NLoS paths.}\label{fig:ROC2}
\end{figure}

Fig. \ref{fig:ROC1} presents the receiver operating characteristic (ROC) curves of $10^5$ independent Monte Carlo simulations. The experimental parameters are listed in Table \ref{table:one}. The RCS reflection parameter is set according to the Swelling-I type.
To demonstrate the advantages of the proposed weighted GLRT detector for multipath exploitation and the associated joint design of transmit power and receive coefficient (``Proposed, joint design''), we also include the scheme only using the transmit power design but with equal detector coefficients (``Proposed, transmit design''), the scheme only using detector coefficient design but with equal transmit power (``Proposed, detection design''), the scheme with equal transmit power and detector coefficients (``Proposed, w./o design''), and the conventional detector that only exploits the LoS path (``Conventional, LoS only'') for comparisons.
We observe from Fig. \ref{fig:ROC1} that the proposed weighted GLRT detector with the associated joint design can achieve significantly better target detection performance than the conventional detector. More importantly, comparing the schemes with/without design at the transmitter/receiver end, the design of the detector weight coefficient plays a crucial role in effectively utilizing multipath diversity to improve target detection performance. This confirms the importance of the proposed weighted GLRT detector. 

To further explore the advancement of the proposed detector with multipath exploitation, in Fig. \ref{fig:ROC2} we analyze the effects of different RCS coefficients of the NLoS paths. For $P_\text{FA} = 10^{-3}$ and a fixed total RCS variance, the probability of detection is plotted as a function of the RCS variance of the NLoS paths. The results indicate that as the RCS variance of the NLoS paths increases, the proposed weighted GLRT detector maintains a consistently high level of detection performance. In contrast, the conventional detector, which relies solely on the LoS path, suffers significant performance degradation due to the reduced strength of the LoS component. These findings demonstrate that leveraging additional observations from NLoS paths enhances spatial diversity, leading to a substantial improvement in detection performance, especially under fluctuating RCS conditions.

\begin{figure}[!t]
\centering
\subfigure[Proposed detector without design.]{
		\includegraphics[width = 0.48\linewidth]{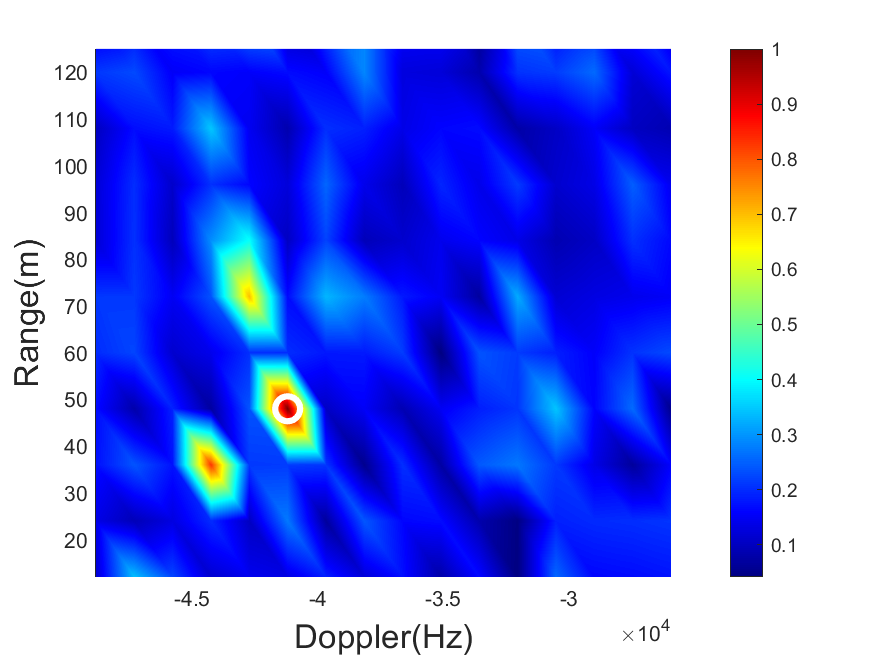}\label{fig:2D map proposed1}}\vspace{-0.2 cm}
\subfigure[Proposed detector with design.]{
    \includegraphics[width = 0.48\linewidth]{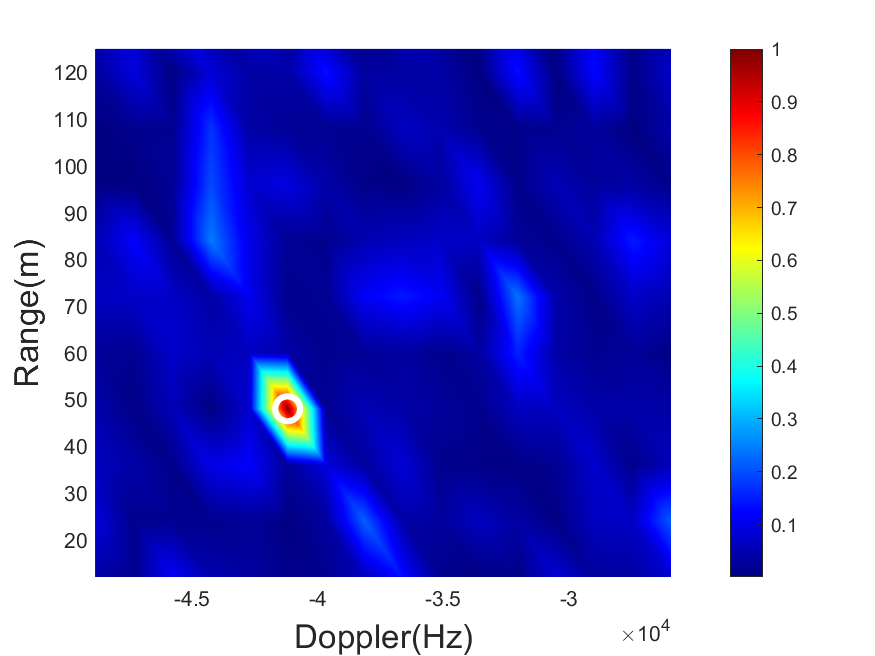}\label{fig:2D map proposed opt}}
\caption{Two-dimensional delay-Doppler results (the target is marked as a white circle).}\label{fig:2D map} \vspace{-0.2 cm}
\end{figure}

Figure \ref{fig:2D map} presents the echo signal processing results across the range-Doppler grid for the area of interest. The detector without joint optimization identifies one true target along with two ghost targets, reflecting its limitations. In contrast, the proposed detector with joint design exhibits more robust detection, accurately identifying the true target without spurious detections.

 \vspace{-0.2 cm}

\vspace{-0.2 cm}
\section{Conclusion}
In this paper, we investigated the potential of multipath exploitation for better target detection performance in an OFDM-ISAC system. We first introduced a weighted GLRT-based detector by leveraging the diversity in the delay-Doppler domain. Then, the joint transmit power allocation and target detection design problem was formulated and solved to maximize the target detection probability subject to the BS power constraint and the SNR constraint of the communication user. Simulation results verified the benefits of properly utilizing multipath signals and the effectiveness of the proposed algorithm.


\vfill


\begin{thebibliography}{99}
\bibliographystyle{IEEEtran}


\bibitem{ISAC summary}
F. Liu, Y. Cui, C. Masouros, J. Xu, T. X. Han, Y. C. Eldar, and S. Buzz, ``Integrated sensing and communications: Toward dual-functional wireless networks for 6G and beyond,'' \textit{IEEE J. Sel. Areas Commun.}, vol. 40, no. 6, pp. 1728-1767, Jun. 2022.
\bibitem{ISAC LR}
R. Liu, M. Li, Y. Liu, Q. Wu, and Q. Liu, ``Joint transmit waveform and passive beamforming design for RIS-aided DFRC systems," \textit{IEEE J. Sel. Topics Signal Process.}, vol. 16, no. 5, pp. 995-1010, Aug. 2022.

\bibitem{sensing assisted comm}
A. Zhang, M. L. Rahman, X. Huang, Y. J. Guo, S. Chen, and R. W. Heath, ``Perceptive mobile networks: Cellular networks with radio vision via joint communication and radar sensing,'' \textit{IEEE Veh. Technol. Mag.}, vol. 16, no. 2, pp. 20-30, Jun. 2021.
%
\bibitem{diss1}
M. Leigsnering, F. Ahmad, M. G. Amin, and A. M. Zoubir, ``Multipath exploitation in sparse scene recovery using sensing-through-wall distributed radar sensor configurations,'' in \textit{Proc. IEEE Int. Conf. Acoust., Speech, Signal Process. (ICASSP)}, South Brisbane, Australia, Apr. 2015, pp. 2749-2753.

\bibitem{diss2}
S. Sen and A. Nehorai, ``Adaptive OFDM radar for target detection in multipath scenarios'', \textit{IEEE Trans. Signal Process.}, vol. 59, no. 1, pp. 78-90, Jan. 2011.
\bibitem{diss3}
Z. Xu, C. Fan, and X. Huang, ``MIMO radar waveform design for multipath exploitation,'' \textit{IEEE Trans. Signal Process.}, vol. 69, pp. 5359-5371, Oct. 2021.
%
\bibitem{diss4}
M. Rihan, E. Grossi, L. Venturino, and S. Buzzi, ``Spatial diversity in radar detection via active reconfigurable intelligent surfaces,'' \textit{IEEE Signal Process. Lett.}, vol. 29, pp. 1242-1246, May 2022.
\bibitem{otfs}
R. Hadani, S. Rakib, M. Tsatsanis, A. Monk, A. J. Goldsmith, A. F. Molisch, and R. Calderbank, ``Orthogonal time frequency space modulation,'' in \textit{Proc. IEEE Wireless Commun. Netw. Conf. (WCNC)}, San Francisco, CA, Apr. 2017, pp. 1-6.

\bibitem{np detector}
S. M. Kay, \textit{Fundamentals of Statistical Signal Processing: Detection Theory}, vol. 2. Englewood Cliffs, NJ, USA: Prentice-Hall, 1998.

\bibitem{glrt rate}
A. Dogandzic and A. Nehorai, ``Generalized multivariate analysis of variance: A unified framework for signal processing in correlated noise,'' \textit{IEEE Signal Process. Mag.}, vol. 20, pp. 39-54, Sep. 2003.

\bibitem{cs theorem}
G. Letac and H. Massam, ``Craig-Sakamoto's theorem for the Wishart distributions on symmetric cones,'' \textit{Ann. Inst. Stat. Math.}, vol. 47, pp. 785-799, Feb. 1995.


\end{thebibliography}
\end{document}